\input{aipcheck}

\documentclass[
    ,final            
  ]
  {aipproc}

\layoutstyle{6x9}

\def\Mjup{M$_{\rm Jup}$}

\begin{document}

\title{The LAOG-Planet Imaging Surveys}


\author{G. Chauvin}{
  address={Laboratoire d'Astrophysique, Observatoire de Grenoble, UJF, CNRS:
        414, Rue de la piscine, 38400 Saint-Martin d'H\`eres, France}}

\author{A.-M. Lagrange}{
  address={Laboratoire d'Astrophysique, Observatoire de Grenoble, UJF, CNRS:
        414, Rue de la piscine, 38400 Saint-Martin d'H\`eres, France}}

\author{D. Mouillet}{
  address={Laboratoire d'Astrophysique, Observatoire de Grenoble, UJF, CNRS:
        414, Rue de la piscine, 38400 Saint-Martin d'H\`eres, France}}

\author{J.-L. Beuzit}{
  address={Laboratoire d'Astrophysique, Observatoire de Grenoble, UJF, CNRS:
        414, Rue de la piscine, 38400 Saint-Martin d'H\`eres, France}}

\author{H. Beust}{
  address={Laboratoire d'Astrophysique, Observatoire de Grenoble, UJF, CNRS:
        414, Rue de la piscine, 38400 Saint-Martin d'H\`eres, France}}

\author{D. Ehrenreich}{
  address={Laboratoire d'Astrophysique, Observatoire de Grenoble, UJF, CNRS:
        414, Rue de la piscine, 38400 Saint-Martin d'H\`eres, France}}

\author{M. Bonnefoy}{
  address={Laboratoire d'Astrophysique, Observatoire de Grenoble, UJF, CNRS:
        414, Rue de la piscine, 38400 Saint-Martin d'H\`eres, France}}

\author{F. Allard}{
  address={Centre de Recherche Astronomique de Lyon, 
  46 all\'ee d'Italie, 69364 Lyon cedex 7, France}}

\author{M. Bessel}{
  address={Research School of Astronomy and Astrophysics Institute of Advance Studies, 
	Australian National University: Cotter Road, Weston Creek, Canberra, ACT 2611, Australia}}

\author{M. Bonavita}{
  address={Universita' di Padova, Dipartimento di Astronomia, Vicolo dell'Osservatorio 2, 35122 Padova, Italy}}

\author{S. Desidera}{
  address={Universita' di Padova, Dipartimento di Astronomia, Vicolo dell'Osservatorio 2, 35122 Padova, Italy}}

\author{C. Dumas}{
  address={European Southern Observatory: Casilla 19001, Santiago 19, Chile}}

\author{J. Farihi}{
  address={Department of Physics \& Astronomy, University of Leicester, Leicester LE1 7RH, United Kingdom}}

\author{T. Fusco}{
  address={Office National d'\'Etudes et de Recherches A\'erospatiales, 
  29 avenue de la Division Leclerc, 92322	Ch\^atillon, France}}

\author{D. Gratadour}{
  address={LESIA, Observatoire de Paris, 
  5 place Jules Janssen, 92195 Meudon, France}}

\author{P. Lowrance}{
  address={Spitzer Science Center, IPAC/Caltech: MS 220-6, Pasadena, CA 91125, USA}}

\author{M. Mayor}{
  address={Observatoire de Gen\`eve, 51 Ch. des Maillettes, 1290 Sauverny, Switzerland}}

\author{D. Rouan}{
  address={LESIA, Observatoire de Paris, 
  5 place Jules Janssen, 92195 Meudon, France}}

\author{I. Song}{
  address={Department of Physics \& Astronomy, University of Georgia, Athens, GA 30602-2451, USA }} 

\author{S. Udry}{
  address={Observatoire de Gen\`eve, 51 Ch. des Maillettes, 1290 Sauverny, Switzerland}}

\author{B. Zuckerman}{
  address={Department of Physics \& Astronomy and Center for Astrobiology, 
	University of California: Los Angeles, Box 951562, CA 90095, USA}}

\begin{abstract}

With the development of high contrast imaging techniques and infrared
detectors, vast efforts have been devoted during the past decade to
detect and characterize lighter, cooler and closer companions to
nearby stars, and ultimately image new planetary
systems. Complementary to other observing techniques (radial velocity,
transit, micro-lensing, pulsar-timing), this approach has opened a new
astrophysical window to study the physical properties and the
formation mechanisms of brown dwarfs and planets. I here will briefly
present the observing challenge, the different observing techniques,
strategies and samples of current exoplanet imaging searches that have
been selected in the context of the LAOG-Planet Imaging Surveys. I
will finally describe the most recent results that led to the
discovery of giant planets probably formed like the ones of our solar
system, offering exciting and attractive perspectives for the future
generation of deep imaging instruments.

\end{abstract}

\maketitle


\section{Searching for planets}

The search for planets has been an important driver for observers in
the two last decades. Their detection and characterization contribute
to developing our understanding of their structure, formation and
evolution.  In the close ($\le 5$~AU) environment of stars, the radial
velocity, transit, micro-lensing, pulsar-timing observing techniques
are so far best suited. The radial velocity (RV) and transit
techniques are nowadays the most successful methods for detecting and
characterizing the properties of exo-planetary systems. The RV surveys
have focused on main sequence solar-type stars, with numerous narrow
optical lines and low activity, to ensure high RV precision. Recently,
planet-search programs have been extended to lower and higher mass
stars (\cite{endl06}, \cite{lagrange09}) and younger and more evolved
systems (\cite{joer05}, \cite{john07}). Since the discovery of 51
Peg\,b (\cite{mayor95}), more than 300 exo-planets have been
identified featuring a broad range of physical (mass) and orbital (P,
$e$) characteristics (\cite{udry07}, \cite{butler06}). This technique
also revealed the existence of the so-called brown dwarf desert at
small ($\le 5$~AU) separations (\cite{gret2006}). The bimodal aspect
of the secondary mass distribution indicates different formation
mechanisms for two populations of substellar companions, brown dwarfs
and planets. The transit technique coupled with RV enables
determination of the radius and density of giant planets and thus a
probe of their internal structure. Moreover, spectral elements of a
planetary atmosphere can be revealed during primary or secondary
eclipse (\cite{swai08}, \cite{gril08}). To extend such systematic
characterization at larger scales ($\ge10$~AU), the deep imaging
technique is particularly well suited to probe the existence of
planets and brown dwarf companions and complete our view of planetary
formation and evolution. To access small angular separations, the
space telescope (HST) or the combination of Adaptive Optics (AO)
system with very large ground-based telescopes (Palomar, CFHT, Keck,
Gemini, Subaru, VLT) have become mandatory. We present here the
results of the 3 main planet search survey conducted at the Grenoble
institute of Astrophysics (LAOG) since 2002.

\section{The environment of exoplanet host stars}

We have conducted a deep coronographic AO imaging survey of 26 stars
with planets detected through radial velocity measurements (\cite{cha06}). The domain
investigated typically ranges between $0.1~\!''$ to $15~\!''$
(i.e. about 3 to 500~AU, according to the mean distance of the
sample). The survey is sensitive to stellar and substellar companions
with masses greater than $30~\rm{M}_{\rm{Jup}}$ (0.5~Gyr) with CFHT
and $15~\rm{M}_{\rm{Jup}}$ (0.5~Gyr) with VLT, at $2~\!''$
($\sim60$~AU) from the primary star. Among the 20 stars that were
found to have potential companions (candidate companions, hereafter
CCs), we could identify (through proper motion measurements) and then
monitor 3 bound companions to three stars: 1) HD1237 is surrounded by
a 0.13 M$_{\odot}$ star at about 70 AU (proj. distance) with an
orbital motion marginally resolved; 2) HD 27442 is surrounded by a
white dwarf located at 240 AU. This is the third white dwarf
discovered sofar around a star hosting planets, in addition to Gl86\,B
and HD\,147513\,B (\cite{pot97}). In both cases, the RV
drift induced by these companions are below the limits of current RV
studies. This illustrates the complementarity between direct imaging
and RV studies. Finally, we also found a 0.55 M$_{\odot}$ star
(assuming an age of 0.5 Gyr) orbiting HD196885. This latter case
brings to 5 the number of stars hosting planets and members of close
in ($\leq$ 20 AU) multiple systems. Combining AO and RV data, we were
able to already strongly constrain the mass and orbit of both systems:
Gl86\,B (0.48-0.54~M$_{\odot}$; $e\ge$0.4; $a = 18$ AU; \cite{lag06} and Fig.~1) and HD\,196885\,B (M=0.55~M$_{\odot}$ ;
$e=0.45$; $a = 26$; \cite{cha07}). We additionaly
traced back the evolutive and dynamical history of Gl86\,B, which was
originally a G dwarf on a close ($a_{ini} = 13$ AU;) and less
eccentric orbit. Several open questions remain related to the fact
that the exoplanet must have survived all the late evolution stages of
Gl86\,B.

\begin{figure}
  \includegraphics[height=.2\textheight]{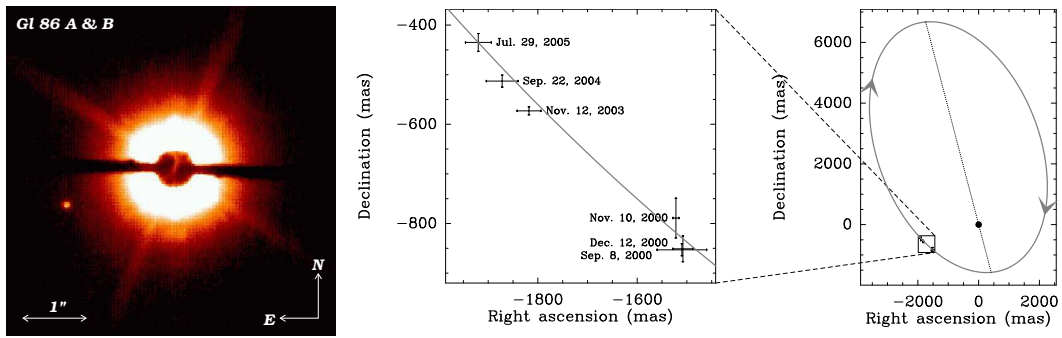}

  \caption{ \textbf{Left}: VLT/NACO Ks-band coronagraphic image
of GLiese 86~A and B, acquired on September 24, 2004, with an occulting mask
of diameter 0.7~$\!''$. \textbf{Right}: A representation of the
orbital solutions, as
projected onto the plane of the sky (view of the full orbit and an
enlargement).}

\end{figure}

\section{Probing planets orbiting dusty stars}

The dusty star survey was focused on probring the presence of planets
around about 15 stars with known debris disks -- disks containing dust
which is not primordial but produced by collisions among larger rocky
bodies. In the course of this survey, a companion candidate was
detected around $\beta$~Pic, a A5V star at a distance of
$19.3\pm0.2$~pc (\cite{crifo97}), which remains the best studied young
($12^{+8}_{-4}$~Myr;
\cite{zuckerman01}) system, with an impressive amount of indirect
signs pointing toward the presence of planets.

The disk shows a relative inner void of matter inside
$50$~AU. \cite{lec95}
presented intriguing light variations possibly due to disk
inhomogeneities produced by a Jupiter size planet at $>6$~AU.
Several asymmetries have been identified in the disk at optical
(\cite{kalas95}, \cite{heap00}) and infrared (\cite{telesco05})
wavelengths, as well as a warp at $\sim 50$~AU\ (\cite{mouillet97};
\cite{heap00}). The structure is well reproduced by the deformation
induced on colliding planetesimals by a giant planet on a slightly
inclined orbit within 50~AU\ from the star (\cite{krist96},
\cite{mouillet97}, \cite{gorka00}, \cite{augereau01} and
\cite{thebault01}). Using deep adaptive-optics $L'$-band images, 
a faint point-like signal is detected at a projected distance of
$\simeq 8$~AU\ from the star, within the North-East side of the dust
disk (see Fig.~2). Various tests were made to rule out with a good confidence level
possible instrumental or atmospheric artefacts. The probability of a
foreground or background contaminant is extremely low, based in
addition on the analysis of previous deep HST images.  Its $L'=11.2$
apparent magnitude would indicate a typical temperature of $\sim
1500$~K and a mass of $\sim8$~\Mjup. If confirmed, it could explain
the main morphological and dynamical peculiarities of the $\beta$~Pic
system. During the same year, \cite{marois08b}) and
\cite{kalas08} have reported images of giant planets to
intermediate-mass stars. The present detection remains unique
by the proximity of the resolved planet. Its closeness and location
inside the $\beta$~Pic disk suggest a formation process by core accretion or
disk instabilities rather than binary like formation processes.

\begin{figure}
  \includegraphics[height=.2\textheight]{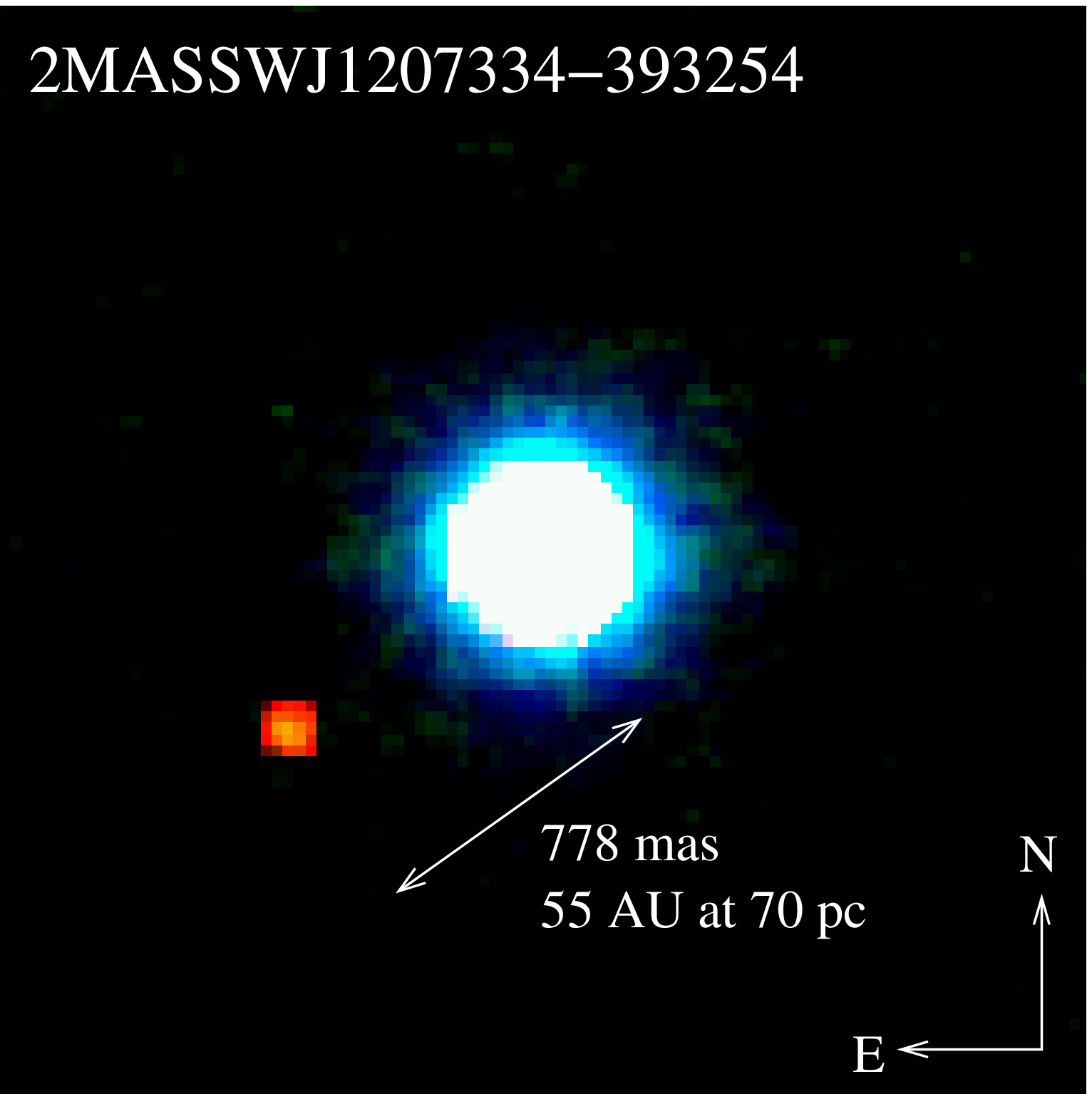}
  \includegraphics[height=.2\textheight]{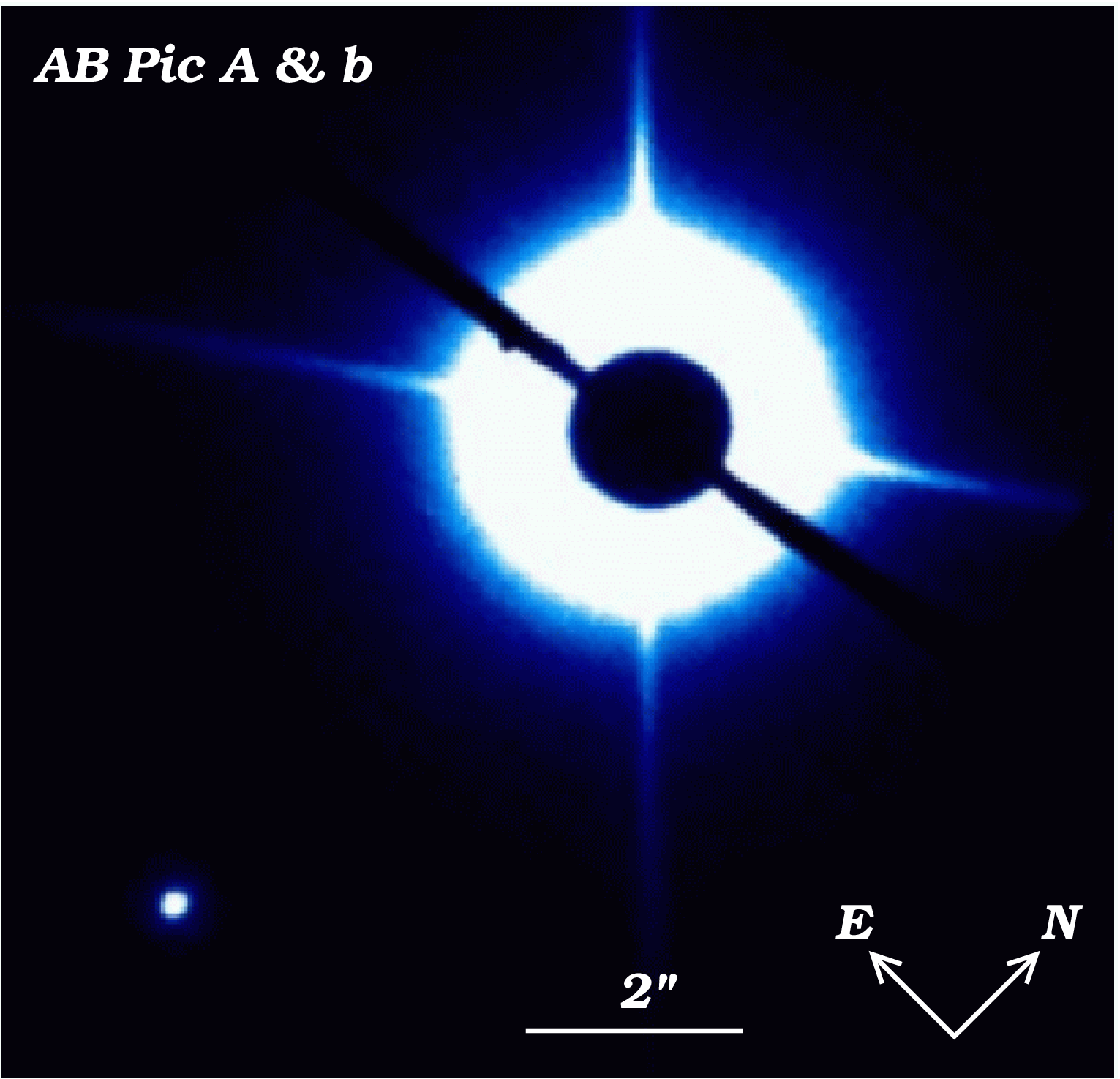}
  \includegraphics[height=.15\textheight]{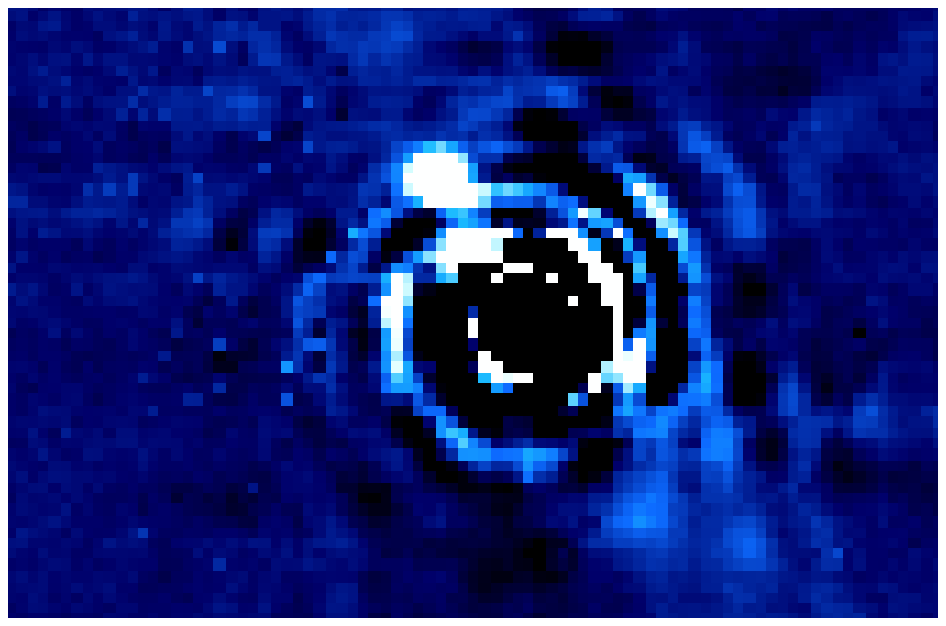}

  \caption{ \textbf{Left}: VLT/NACO deep image of the TWA
brown dwarf 2M1207 and its 5~M$_{\rm{Jup}}$ companion at 41~AU
obtained in $K_s$-band with the S27 camera (Chauvin et
al. 2004). \textbf{Middle}: VLT/NACO coronagraphic imaging with the
small ($\oslash = 0.7~\!''$) coronagraphic mask of the
13~M$_{\rm{Jup}}$ companion to the AB Pic star, member of the Tuc-Hor
association (\cite{cha05b}). \textbf{Right}: VLT/NACO deep
L~$\!'$ image of the $\beta$ Pictoris b candidate detected thanks to
an optimized PSF-reference subtraction.}
\end{figure}

\section{Young, nearby austral stars}

Finally, we have conducted a deep adaptive optics imaging survey with NACO at
the VLT of 88 nearby stars of the southern hemisphere (\cite{cha09}).  Our selection
criteria favored youth ($\le100$~Myr) and proximity to Earth
($\le100$~pc) to optimize the detection of close planetary mass
companions. Known visual binaries were excluded to avoid degrading the
NACO AO and/or coronagraphic detection performances. Among our sample,
51 stars are members of young, nearby comoving groups. 32 are young,
nearby stars currently not identified as members of any currently
known association and 5 have been reclassified as older ($\ge100$~Myr)
systems. The spectral types cover the sequence from B to M spectral
types with $19\%$ BAF stars, $48\%$ GK stars and $33\%$ M dwarfs.  The
separation investigated typically ranges between $0.1~\!''$ to
$10~\!''$, i.e. between typically 10 to 500~AU.  A sample of 65 stars
was observed in deep coronagraphic imaging to enhance our contrast
performances to $10^{-6}$ and to be sensitive to planetary mass
companions down to 1~\Mjup\ (at 24\% of our sample) and 3~\Mjup\ (at
67\%).  We used a standard observing sequence to precisely measure the
position and the flux of all detected sources relative to their visual
primary star. Repeated observations at several epochs enabled us to
discriminate comoving companions from background objects. In the
course of that survey, we discovered of 17 new close ($0.1-5.0~\,''$)
multiple systems. HIP\,108195\,AB and C (F1III-M6), HIP\,84642\,AB
($a\sim14$~AU, K0-M5) and TWA22\,AB ($a\sim1.8$~AU; M6-M6) are
confirmed as comoving systems. TWA22\,AB, with 80\% of its orbit
already resolved, is likely to be a rare astrometric calibrator for
testing evolutionary model predictions. About $236$ faint CCs were
detected around 36 stars observed in coronagraphy. Follow-up
observations with VLT or HST for 30 stars enabled us to identify their
status. 1\% of the CCs detected have been confirmed as comoving
companions, 43\% have been identified as probable background
contaminants and about 56\% need further follow-up observations. The
remaining CCs come mostly from the presence of crowded fields in the
background of the 6 stars observed at one epoch. We confirmed
previously discovered substellar companions around GSC\,08047-00232 (\cite{cha03}, \cite{cha05a}),
AB\,Pic (\cite{cha05b}) and 2M1207 (\cite{cha04}, \cite{cha05a}) and placed them in the perspective of confirmed
substellar companions among young, nearby associations (see Fig.~3). Finally, the
statistical analysis of our complete set of detection limits enables
us to constrain at large semi-major axes, 20 to a few 100 AU, various
mass, period and eccentricity distributions of giant planets
extrapolated and normalized from RV surveys. It enables us to derive
limits on the occurence of giant planets for a given set of physical
and orbital distributions.  The survey starts constraining
significantly the population of giant planet for masses $\ge3$~\Mjup (see Fig.~4).

\begin{figure}
  \includegraphics[height=.35\textheight]{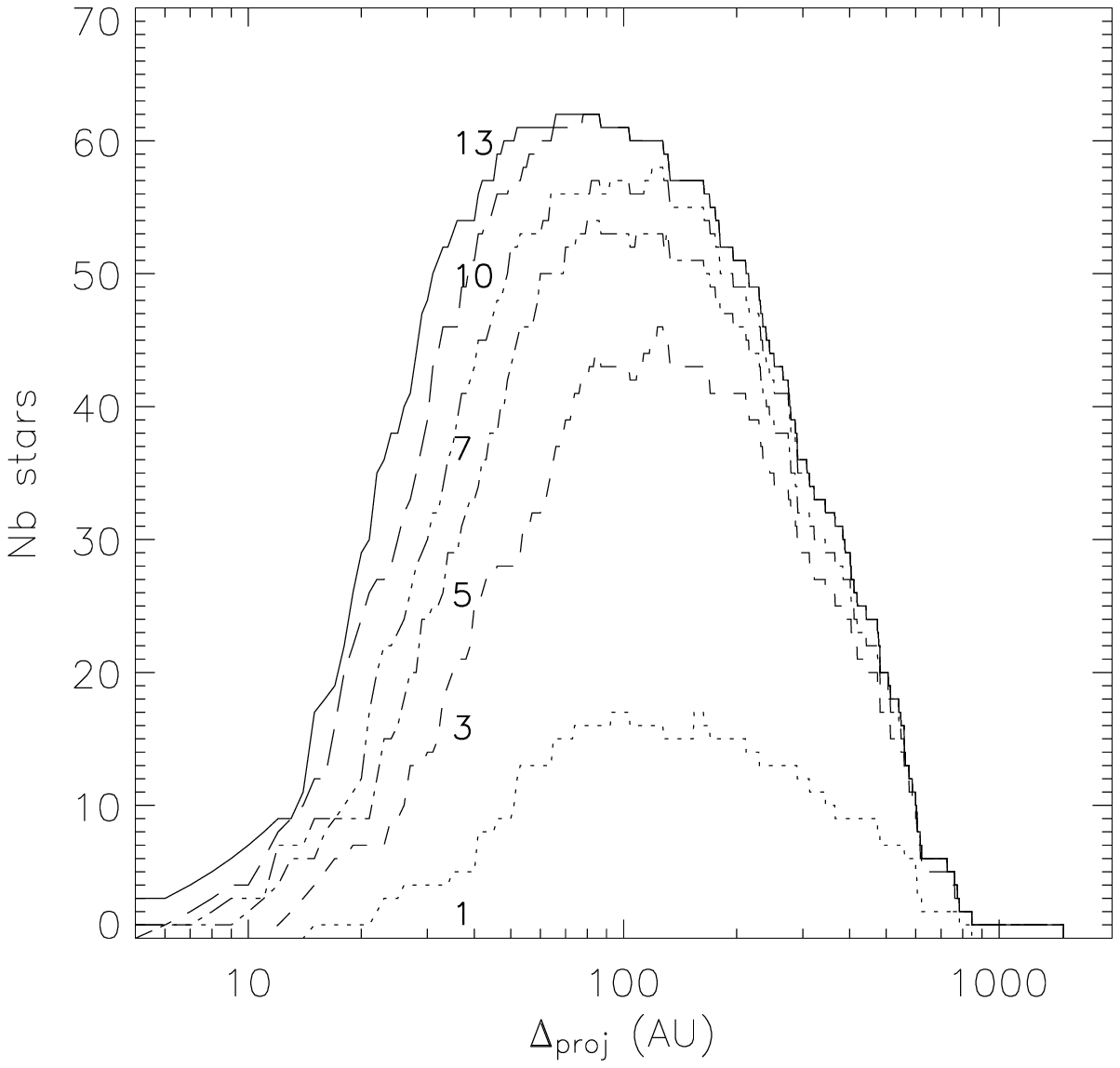}
  \includegraphics[height=.35\textheight]{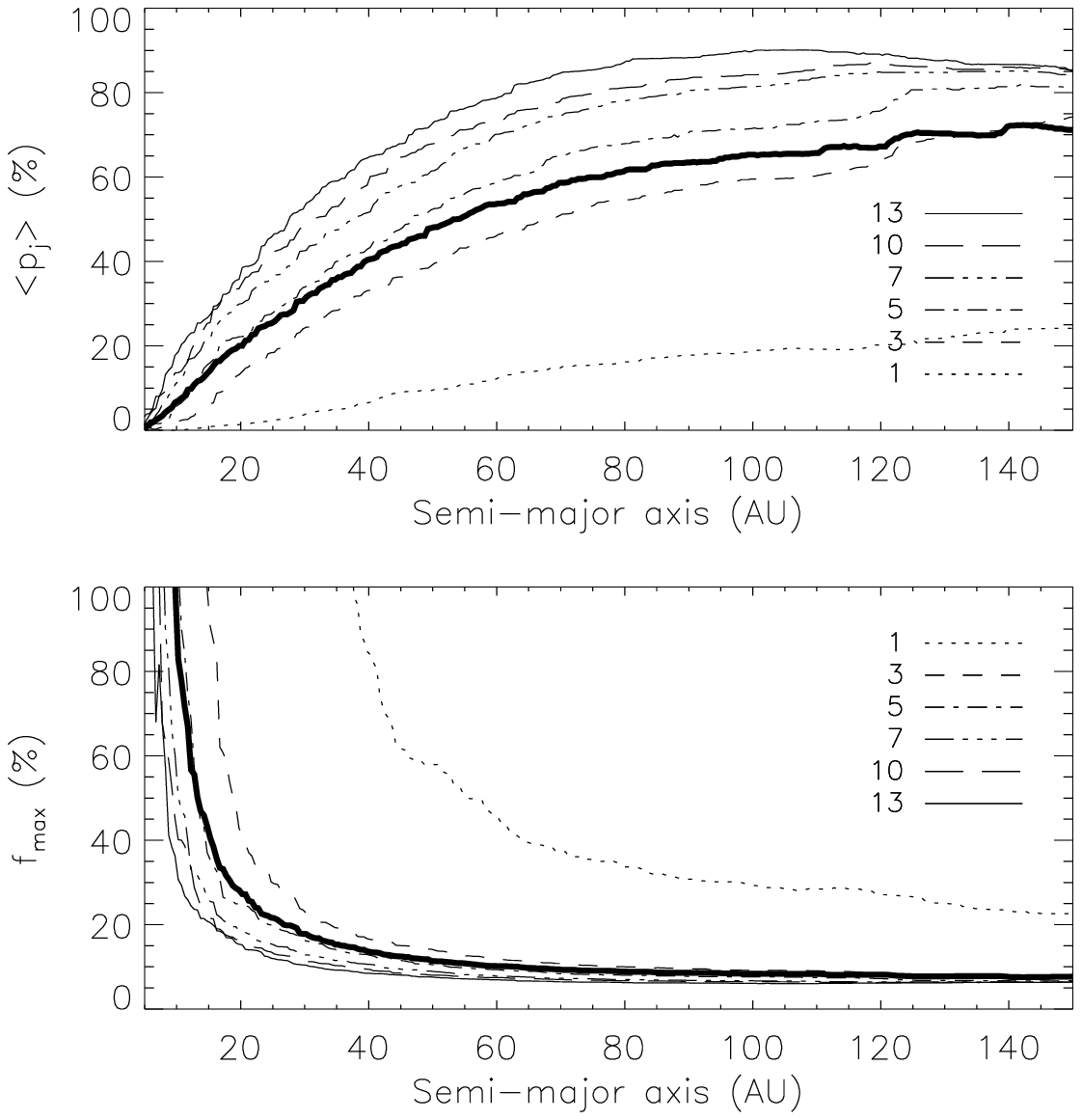}

  \caption{\textbf{Left:} Histogram of projected physical separations explored, for
various planetary masses (1, 3, 5, 7, 10 and 13)~M$_{\rm{Jup}}$, in
the close vicinity of the 65 young, nearby stars observed with NACO at
VLT in coronagraphy (\cite{cha09}). Contrast performances have been converted into
masses based on the nIR photometry, age and distance of the primary
stars.  \textbf{Right:}, \textit{Top}, Survey mean detection probability derived as a
function of semi-major axis assuming parametric mass and period
distributions derived by \cite{cumm08}, i.e with
$\alpha=-1.31$, $\beta=-0.74$ and $\gamma=1.25$. The results are
reported for individual masses: 1, 3, 5, 7, 10 and 13~\Mjup. The
integrated probability for the planetary mass regime is shown with the
\textit{thick solid} line. \textit{Bottom}, Planet fraction upper
limit derived as a function of semi-major axis, given the same
mass and period distributions.}
\end{figure}

\section{Conclusion and perspective}

In the first few years following the discovery of the companion to
2M1207 (\cite{cha04}, all planetary mass companions were
discovered at relatively wide separations or with small mass ratio
with their primaries. However, the recent discoveries of planetary
mass objects around the star Fomalhaut (\cite{kalas08}), HR\,8799
(\cite{marois08b}) and $\beta$ Pictoris (Lagrange et al. 2008), now
open a new era for the deep imaging study of giant planets that
probably formed like those of our solar system. In the perspective of
on-going and future deep imaging instruments either from the ground
(Gemini/NICI, Subaru/HiCIAO, SPHERE, GPI, EPICS) or from space (JWST,
TPF/Darwin), this work represents a pioneer successful study,
providing, with other surveys, precise information (stellar and
substellar multiplicity, non-detections and background contaminants)
to better characterize the overall environment of young, nearby stars,
that will be prime targets for futur exoplanets search.




\begin{theacknowledgments}

We thank all the organization (LOC and SOC) of the 2nd Subaru
International Conference Disks to Exoplanets. For all results
presentes, we thank the ESO Paranal staff for performing the service
mode observations. We also acknowledge partial financial support from
the PNPS and Agence National de la Recherche, in France, from INAF
through PRIN 2006 ``From disk to planetray systems: understanding the
origin and demographics of solar and extrasolar planetary systems''
and from NASA in the USA.

\end{theacknowledgments}



\bibliographystyle{aipproc}   



\end{document}